\documentclass[lettersize,journal]{IEEEtran}
\usepackage{amsmath,amsfonts}
\usepackage{algorithmic}
\usepackage{float}
\usepackage{booktabs}
\usepackage{multirow}
\usepackage{algorithm}
\usepackage{subfigure} 
\usepackage{array}
\usepackage{multirow}
\usepackage[caption=false,font=normalsize,labelfont=sf,textfont=sf]{subfig}
\usepackage{textcomp}
\usepackage{stfloats}
\usepackage{url}
\usepackage{verbatim}
\usepackage{graphicx}
\usepackage{cite}
\usepackage[T1]{fontenc}
\usepackage{etoolbox}
\makeatletter
\patchcmd{\@makecaption}
{\scshape}
{}
{}
{}
\makeatother
\begin{document}

\title{GraphMU: Repairing Robustness of Graph Neural Networks via Machine Unlearning}
\author{Tao~Wu, Xinwen Cao, Chao Wang, Shaojie Qiao, Xingping Xian, Lin Yuan, Canyixing Cui, Yanbing Liu
	
	\thanks{T. Wu, X. Cao, X. Xian, L. Yuan, C. Cui, and Y. Liu work at the School of Cybersecurity and Information Law, Chongqing University of Posts and Telecommunications, Chongqing 400065, China. S. Qiao is a Professor with the School of Software Engineering, Chengdu University of Information Technology, Chengdu 610225, China. C. Wang is an Associate Professor with the School of Computer and Information Science, Chongqing Normal University, Chongqing 401331, China, and Yunnan Key Laboratory of Software Engineering, Yunnan 650504, China. Corresponding author: Xingping~Xian, E-mail: xxp0213@gmail.com, Corresponding author: Lin Yuan, E-mail: yuanlin@cqupt.edu.cn.
		
		Copyright (c) 2024 IEEE. Personal use of this material is permitted. However, permission to use this material for any other purposes must be obtained from the IEEE by sending a request to pubs-permissions@ieee.org.
}}

\markboth{Journal of \LaTeX\ Class Files,~Vol.~18, No.~9, September~2020}%
{How to Use the IEEEtran \LaTeX \ Templates}

\maketitle

\begin{abstract}
Graph Neural Networks (GNNs) have demonstrated significant application potential in various fields. However, GNNs are still vulnerable to adversarial attacks. Numerous adversarial defense methods on GNNs are proposed to address the problem of adversarial attacks. However, these methods can only serve as a defense before poisoning, but cannot repair poisoned GNN. Therefore, there is an urgent need for a method to repair poisoned GNN. In this paper, we address this gap by introducing the novel concept of model repair for GNNs. We propose a repair framework, Repairing Robustness of Graph Neural Networks via Machine Unlearning (GraphMU), which aims to fine-tune poisoned GNN to forget adversarial samples without the need for complete retraining. We also introduce a unlearning validation method to ensure that our approach effectively forget specified poisoned data. To evaluate the effectiveness of GraphMU, we explore three fine-tuned subgraph construction scenarios based on the available perturbation information: ({\romannumeral1}) Known Perturbation Ratios, ({\romannumeral2}) Known Complete Knowledge of Perturbations, and ({\romannumeral3}) Unknown any Knowledge of Perturbations. Our extensive experiments, conducted across four citation datasets and four adversarial attack scenarios, demonstrate that GraphMU can effectively restore the performance of poisoned GNN.
\end{abstract}


\section{Introduction}
\IEEEPARstart{G}{raph} neural networks (GNNs) have emerged as a powerful tool for processing graph-structured data, in which each node updates its hidden representation by aggregating information from neighborhoods. In recent years, GNNs have found widespread usage and achieved state-of-the-art performance in various domains. Despite the success, concerns have been raised regarding the robustness of GNNs when subjected to adversarial attacks \cite{zugner2018adversarial, zhou2021hierarchical, lin2023exploratory}. Such attacks generate adversarial examples by modifying graphs with imperceptible perturbations, so as to substantially degrade the performance of GNNs. The phenomenon has posed great challenges to apply GNNs in safety-critical applications such as financial systems, risk management and medical diagnoses.

To defend against adversarial attacks, considerable attempts have been made, which can be divided into three categories: graph adversarial training, robust model design and graph purification. Specifically, ({\romannumeral1}) Graph adversarial training aims to enhance the robustness of GNNs by incorporating adversarial examples into the training procedure. The methods include Spectral Adversarial Training (SAT) \cite{li2022spectral}, Directional Graph Adversarial Training (DGAT) \cite{hu2021robust} and Adversarially Regularized Variational Graph Autoencoder (ARVGA) \cite{pan2019learning}, etc. ({\romannumeral2}) Robust model design focuses on optimizing the architecture of GNNs to withstand adversarial perturbations. Examples of such methods include Penalized Aggregation GNN (PA-GNN) \cite{tang2020transferring}, Robust GCN (RGCN) \cite{zhu2019robust}, Similarity-based GNN (GSP-GNN)\cite{yao2023defending}, etc. ({\romannumeral3}) Graph purification involves the detection and removal of potential adversarial perturbations from input graphs before the model is trained. The methods in this type of research include Vaccinated-GCN \cite{entezari2020all}, Heterogeneous Graph Purification Network (HGPN) \cite{shen2023heterogeneous}, Federated Framework with Joint Graph Purification (FedGP) \cite{chen2023medical}, GNNs with Global Noise Filtering (GNN-GNF) \cite{feng2022graph}, etc. 
\begin{figure}[!t]
	\centering
	\includegraphics[width=3.7in]{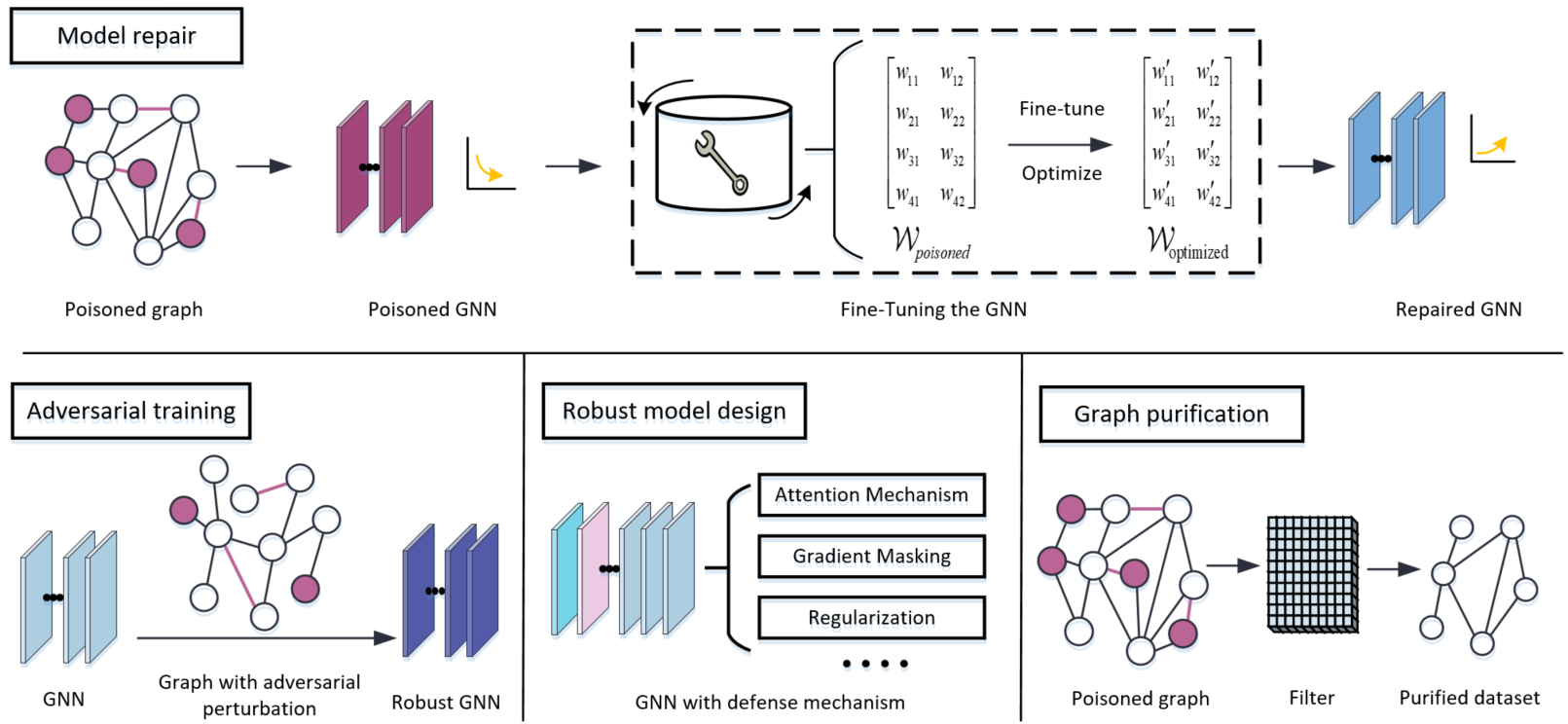}
	\caption{Illustration of model repair framework and its difference from traditional adversarial defense methods adversarial training, robust model design, and graph purification. }
	\label{defense}
\end{figure} 

Although the aforementioned methods have been proposed to defend against adversarial attacks, they all employ static defense strategies that are unable to adapt to time-varying intelligent application scenarios of GNNs. In particular, the graph adversarial training methods are only effective against known adversarial attacks, and the unknown adversarial examples contained in the input data of the deployed GNNs still lead to a significant drop in their performance. Robust model design often targets specific flaws of GNNs and is based on incomplete and not completely accurate model of the world. The resulting models cannot adapt to newly emerging defects and newly arrived knowledge. Moreover, graph purification methods are often based on specific prior assumption about the pattern of graph data. Given the above, after the deployment of GNNs models, the only way to compensate for newly discovered defects is to retrain the models. Meanwhile, due to the uncertainty of the emergence of model defects, multiple model retrainings may be required for defect repair. However, retraining GNNs from scratch requires a large number of computational resources, data samples and time overhead, which is not a feasible solution against adversarial attacks. For example, due to the unavailability of the training data used for the deployed GNNs, it is often not feasible to perform a complete retraining of the models. Therefore, it is crucial to design a dynamic model robustness repair framework, as shown in Fig.\ref{defense}, which aims to enhance the adversarial robustness of GNNs at runtime against continuously emerging defects while keeping the modifications minimal.

In this paper, we focus on the defense against poisoning attacks. Usually, the poisoning attacks are performed at train time by injecting a small amount of carefully crafted samples, i.e., adversarial samples, into the training data to manipulate the learned ML model so that its test accuracy decreases. Consequently, the dynamic model robustness repair framework against poisoning attacks should detect the adversarial samples in the training data and eliminate their influence without fully retraining models. Meanwhile, in recent years, machine unlearning \cite{xu2024machine, bourtoule2021machine} reveal that private information about users can be removed from trained machine learning models and the model owner would be shielded from constant and expensive retraining exercises. That is, machine unlearning enables model owners to completely remove the traces of samples that need to be deleted from a trained model, while maintaining the contributions of other samples unaffected, while the computational cost of the unlearning process is significantly lower compared to retraining the model from scratch \cite{wang2023inductive}. Therefore, machine unlearning can be utilized for the defense against poisoning attacks and the dynamic restoration of model robustness.

Despite numerous studies on machine unlearning, unlearning methods on graph data. i.e., graph unlearning, are still lacking due to the node dependency within graphs. Specifically, based on the Sharding, Isolation, Slicing, and Aggregation (SISA) framework \cite{xu2024machine} for exact unlearning, Chen et al. \cite{chen2022graph} proposed GraphEraser, a novel graph unlearning method based on balanced graph partition. Wang et al. \cite{wang2023inductive} proposed a Guided Inductive Graph Unlearning framework (GUIDE) and presented a guided graph partitioning component. However, graph partitioning will result in loosing the global structural information of graphs and ignoring the edges that span the subgraphs, which could hurt the learned node representations. In this paper, our emphasis is on approximate unlearning that aims to minimize the influence of unlearned samples to an acceptable level while achieving an efficient unlearning process. However, existing approximate graph unlearning methods \cite{chien2022certified, wu2023certified, cong2023efficiently} are always model-specific. Thus, we can conclude that the main challenge is to build a transparent and universal unlearning mechanism for various GNN models.

\textbf{Main contributions}. Motivated by our above findings, in this paper we propose the first robustness dynamic repair framework of GNNs based on approximate unlearning called Repairing Robustness of Graph Neural Networks via Machine Unlearning (GraphMU). Briefly, GraphMU comprises three steps: perturbation localization, subgraph construction, and model fine-tuning. Specifically, based on identified adversarial perturbations, it constructs clean subgraphs that have eliminated the influence of adversarial perturbations based on their neighboring subgraphs. Then, it fine-tunes the defective GNNs based on the clean subgraphs. Moreover, depending on the situations of poisoning attacks, GraphMU considers three scenarios: a known proportion of adversarial perturbations (gray-box), complete knowledge of adversarial perturbations (white-box), and complete unawareness of adversarial perturbations (black-box). GraphMU takes into account three types of adversarial perturbations: node injection, structural perturbation, and feature perturbation. To detect these perturbations, GraphMU proposes the use of BWGNN, Jaccard, and SimRank methods, respectively. Additionally, to evaluate the effectiveness of the proposed dynamic repair mechanism, we also introduce a verification mechanism that considers the intrinsic dependencies within graph data.

The main contributions of this study are as follows:
\begin{itemize}
	\item We design a transparent and model-agnostic dynamic repair framework for enhancing the robustness of GNNs on the basis of machine unlearning. This framework can dynamically fine-tune the target GNNs based on identified adversarial perturbations to continuously improve it, which is crucial for the practical deployment of GNNs.
	
	\item We design detection mechanisms for adversarial perturbations and methods for constructing clean subgraphs under different perturbation types and prior knowledge, thereby supporting the fine-tuning of the target GNNs.
	
	\item We design a verification mechanism for evaluating the model robustness repair framework, taking into account the data dependency within graphs, in order to eliminate the impact of identified adversarial perturbations. 
	
	\item We conduct a systematic experimental analysis on the effectiveness of the proposed framework GraphMU, verifying its advantages from multiple perspectives such as time consumption and computing resources.
\end{itemize}

The rest of this paper is organized as follows: Section \uppercase\expandafter{\romannumeral2} discusses related works. Section \uppercase\expandafter{\romannumeral3} presents preliminaries and the problem definitions. We then elaborate the proposed repair method in Section \uppercase\expandafter{\romannumeral4}. Section \uppercase\expandafter{\romannumeral5} introduces our proposed validation method to verify the effectiveness of the proposed method to forget the poisoned samples. Section \uppercase\expandafter{\romannumeral6} compares related methods as well as analyzes the complexity of the proposed repair method. Experimental setup and analysis of results are provided in Section \uppercase\expandafter{\romannumeral7}. Eventually, we conclude this paper and give future directions in Section \uppercase\expandafter{\romannumeral8}.

\section{Related Works}
\subsection{Adversarial Defense on GNNs}
With the rise of adversarial attack on GNNs, corresponding defense methods of GNNs have emerged, and they defend against adversarial attacks from different perspectives. In order to reinforce the robustness of GNNs, Li et al.\cite{li2022spectral} proposed an adversarial training method, SAT, which combines adversarial training and regularization strategies with low-rank approximation and adversarial perturbation of graph structures in the spectral domain. Combining the coding capabilities of graph convolutional networks (GCN), Pan et al.\cite{pan2019learning} proposed a novel adversarial training method. From the perspective of improving the resilience of GNNs against adversarial attacks, Tang et al.\cite{tang2020transferring} design a robust GNN based on penalized aggregation mechanism, PA-GNN, which combines a clean graph and a meta-optimization algorithm to detect adversarial edges and reduce the effects of those. To address the noise problem in heterogeneous graphs, Zhang et al.\cite{shen2023heterogeneous} proposed HGPN, leveraging subgraph-based aggregation and semantic purification for edge refinement. Feng et al.\cite{feng2022graph} propose GNN-GNF for the noise problem in session recommender systems, by introducing an attention mechanism and a global noise filtering module based on GNNs to identify and reduce noise in session data, thereby enhancing the accuracy and robustness of the recommender system.

\subsection{Graph Unlearning}
In recent years, a number of studies have been conducted on graph unlearning, resulting in a wide range of unlearning methods. These methods can be broadly classified into two categories: model-agnostic unlearning and model-intrinsic unlearning. On the one hand, Grapheraser\cite{chen2022graph} and SAFE\cite{dukler2023safe} are well-known model-agnostic unlearning methods based on SISA\cite{bourtoule2021machine} in the field of text and images that employ data partitioning to achieve efficient unlearning without retraining. They divide the dataset into multiple fragments and train a submodel for each fragment. The final prediction is derived by aggregating the predictions of all the sub-models. When data needs to be removed, only the corresponding sub-models need to be retrained partially. However, these methods will weaken the performance of model and have higher computational costs compared to a normally trained model. On the other hand, these model-intrinsic unlearning methods, such as CEU\cite{wu2023certified}, GNNDelete\cite{cheng2022gnndelete} and GraphEditor\cite{cong2022grapheditor}, achieve computational efficiency by adjusting the weights of the pre-trained model to meet specific unlearning criteria, but they lack transparency and are not suitable for poisoned GNNs.

\subsection{Fine-tuning}
Fine-tuning is the process of adjusting a pre-trained model to a specific task with limited labeled data. It can be categorized into several strategies in the field of GNNs: ({\romannumeral1})Direct fine-tuning initializes the pre-trained model's parameters for downstream tasks without updates. For instance, L2P-GNN\cite{lu2021learning} integrates self-supervised learning to mimic fine-tuning during pre-training, enhancing downstream task adaptability. G-TUNING\cite{sun2024fine} preserves generative patterns of downstream graphs by reconstructing graphons, improving transferability across datasets. ({\romannumeral2})Regularized fine-tuning applies additional regularization to maintain model generalizability and prevent overfitting. Notable contributions in this area include GTOT regularizer proposed by Zhang et al.\cite{zhang2022fine} and on- and off-manifold regularization introduced by Kong et al.\cite{kong2020calibrated} which further refine fine-tuning by maintaining local information and calibrating for both in-distribution and out-of-distribution data. ({\romannumeral3})Adapter fine-tuning inserts small, task-specific modules into the pre-trained model for efficient adaptation with minimal parameter updates. Representative work in this strategy includes the innovative approach by Ming et al.\cite{ming2024does}, where they apply adapter fine-tuning to vision-language models, optimizing internal feature representations to bolster in-distribution classification and out-of-distribution detection.

\section{Problem Definition and Preliminaries}
\subsection{Preliminaries}
\paragraph{Notation}Given a graph ${\cal G} = ({\cal A},{\cal X})$ with $n$ nodes, where ${\cal A}_{n{\times}n}\in \{0,1\}$ is the adjacency matrix and ${\cal X}_{n{\times}d}\in \{0,1\}$ is the node features. Let ${\cal V} = \{ {v_i}\} _{i = 1}^n$ represents the set of nodes and ${\cal E} = \{e_{ij}\}$ represents the set of edges where $i,j \in {\cal V}$.

\paragraph{Graph Convolutional Network(GCN)}GCN is one of the classical models in graph-based learning, which has broadly used in many tasks. The convolutional layer and layer-wise propagation rule are defined as:
\begin{equation}
	\begin{array}{l}
		{\cal H}^{(l+1)}=\sigma\left(\tilde{\cal D}^{-\frac{1}{2}} \tilde{\cal A} \tilde{\cal D}^{-\frac{1}{2}} {\cal H}^{(l)} {\cal W}^{(l)}\right),
	\end{array}
	\label{GCN}
\end{equation}
where ${\tilde{\cal A}} = A + {\cal I}_n$ is the adjacency matrix with self-loops. ${\cal I}_n$ is an identity matrix, and ${\tilde{\cal D}}_{ii} = \sum\nolimits_j {{{\tilde{\cal A}}_{ij}}}$ is a diagonal degree matrix. ${\cal W}^{(l)}$ is a trainable input-to-hidden weight matrix. ${\cal H}^{(l)}$ is the matrix of hidden representation(activation). Particularly, ${\cal H}^{(0)} = {\cal X}$ is the input of neural network and ${\cal H}^{(l+1)}$ is the update node feature matrix at layer $l+1$. $\sigma ( \cdot )$ represents the element-wise activation function of network and is usually defined as $ReLU ( \cdot ) = max( \cdot ,0)$.

\paragraph{Target of GNN Attackers}
The goal of the attacker is to seek for a poisoned graph ${\hat {\cal G}}=({\hat {\cal A}},{\hat {\cal X}})$ that can minimize the attack objective ${\cal L}_{atk}$,
\begin{equation}
	\begin{array}{l}
		{max_{\hat {\cal G}}}{{\rm{{\cal L}}}_{atk{\rm{ }}}}\left( {{f_\theta }(\hat {\cal G})} \right) = \sum\limits_{{v_i} \in {{\cal V}}} {{\ell _{atk{\rm{ }}}}} \left( {{f_{{\theta ^*}}}{{(\hat {\cal G})}_{v_i}},{y_{v_i}}} \right),\\
		s.t.,{\rm{ }}\quad {\theta ^*} = \arg {\min _\theta }{{\rm{{\cal L}}}_{train{\rm{ }}}}\left( {{f_\theta }(\hat {\cal G})} \right),
	\end{array}
	\label{GCN}
\end{equation}
where ${{f_\theta }(\hat {\cal G})}$ denotes the prediction of $\hat {\cal G}$ and $\ell ( \cdot , \cdot )$ is a loss function such as cross entropy. ${{\ell _{atk{\rm{ }}}}}$ is a loss function for attacking and one option is to choose ${{\ell _{atk{\rm{ }}}}} = -{\ell}$, and $y_{v_i}$ represent the real label of $v_i$. Note that $\hat {\cal G}$ is constrained to a fixed perturbation budget $\Delta$:
\begin{equation}
	\begin{array}{l}
		\parallel \hat {\cal A} - {\cal A}\parallel  + \parallel \hat {\cal X} - {\cal X}\parallel  \le \Delta.
	\end{array}
	\label{GCN}
\end{equation}

\begin{figure*}
	\centering
	\includegraphics[width=1.0\textwidth]{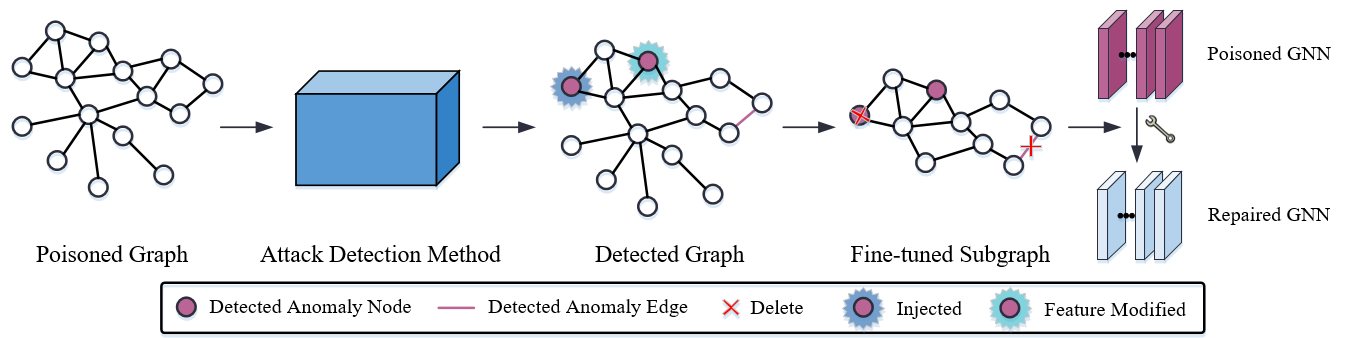}
	\caption{Illustration of GraphMU. Firstly, we use attack detection method to obtain anomalous nodes or edges. According to distinct scenarios of poisoned sample awareness, access to anomalous nodes is also different. Secondly, we construct fine-tuned subgraph based on these detected anomalous nodes or edges. Finally, we use constructed fine-tuned subgraph to optimize the parameters of the poisoned GNN.}
	\label{framework}
\end{figure*}

\subsection{Problem Definition}
\paragraph{Unlearning fine-tuning}Fine-tuning is a common deep learning technique that involves additional training on top of a pre-trained model to fit a specific task or dataset. Traditionally, the whole process of fine-tuning in GNNs can be formulated as follows:
\begin{equation}
	\label{val1}
	\hat \theta=\arg \min _{\theta '} {\cal L}\left(\theta ' ;{\cal D}'\right)+\lambda {\cal R}(\theta '),
\end{equation}
where $\hat \theta$ is the fine-tuned GNN parameters, $L\left(\theta ';{\cal D}'\right)$ which is the loss function for the target task, is used to measure the difference between the prediction of the GNN and the actual label on the target task, $\theta$ denotes the pre-train GNN parameters, $\theta '$ is obtained from $\theta$ with some adjustments, $\lambda$ is the regularization coefficient, and ${\cal R}(\theta)$ is the regularization term. ${\cal D}'=\left\{\left(x_{i}, y_{i}\right)\right\}_{i=1}^{n}$, which denotes datasets for target tasks, here $x_{i}$ is the input example and $y_{i}$ is the corresponding label. Generally, $\cal D$ which is used for pre-training and $\cal D'$ can or can not have the same distribution, but they will not be similar. It can be formally defined as:
\begin{equation}
	\label{val2}
	{\cal D} \sim {{\cal P}_1},{\cal D}' \sim {{\cal P}_2},{\cal S}({\cal D},{\cal D}') < \gamma,
\end{equation}
where both ${\cal P}_1$ and ${\cal P}_2$ denote a certain kind of probability distribution respectively, which can be the same but can be different. ${\cal S}({\cal D},{\cal D}')$ denotes a similarity measure of $\cal D$ and ${\cal D}'$, such as Jaccard similarity, KL scatter. If ${\cal S}({\cal D},{\cal D}') > \gamma$, $\cal D$ and ${\cal D}'$ are considered similar; otherwise they are not similar. In the meantime, $\theta$ need to be adjusted before pre-tuning, or the parameters of certain layers in GNN need to be frozen to adapt to new downstream tasks.

In contrast, in unlearning fine-tuning, ${\cal D}'$ is artificially constructed based on $\cal D$. Moreover, due to ${\cal D}' \subset {\cal D}$, $\theta$ does not need to be adjusted prior to fine-tuning. At the same time, during pre-training, the parameters of each layer of the GNN are obtained based on $\cal D$, so it is necessary to optimize the parameters of each layer of the GNN without freezing the parameters of certain layers of the GNN. The process of unlearning fine-tuning can be formalized as:
\begin{equation}
	\label{val3}
	\hat \theta=\arg \min _{\theta} L\left(\theta  ;{\cal D}'\right).
\end{equation}

The purpose of unlearning fine-tuning is to make the pre-trained model (i.e., the poisoned model) forget poisoned samples. The dataset used for fine-tuning is also derived from the dataset used in the pre-training phase and the downstream tasks are all the same, so there is no need to modify the parameters of GNN prior to fine-tuning.

\section{Methodology}
In this section, we introduce our proposed repair method, GraphMU, as shown in Fig. \ref{framework}. We begin with a concise overview of GraphMU, followed by a detailed explanation of constructing a fine-tuned subgraph based on available perturbation knowledge. We conclude with the process of updating the parameters of affected GNNs with this fine-tuned subgraph.

\subsection{Overview}
The GraphMU framework, shown in Figure \ref{framework}, is designed to weaken the impact of adversarial attack on poisoned GNN through a strategy that refines subgraphs and tunes the network parameters. It addresses three types of adversarial perturbations: node injection, feature modification, and edge addition. The approach adapts based on the available knowledge of these perturbations: known perturbation ratios, complete knowledge of perturbations, and no knowledge of perturbations.

The framework consists of the following key steps: ({\romannumeral1})Detection and Isolation of Adversarial Perturbations: Depending on the scenario, detect anomalies using appropriate algorithms (BWGNN\cite{tang2022rethinking}, Jaccard, SimRank) or directly identify the perturbations when complete information is available; ({\romannumeral2})Construction of Fine-Tuned Subgraphs: For each detected or identified adversarial perturbation, construct subgraphs that exclude the adversarial elements and include the surrounding graph structure. This step is crucial as it forms the basis for the fine-tuning process; ({\romannumeral3})Fine-Tuning of Poisoned GNNs: Utilize the constructed fine-tuned subgraphs to adjust the parameters of the poisoned GNNs. This process incorporates the phase of parameter optimization, wherein the model is trained to mitigate the influence of adversarial inputs and to concentrate on the intrinsic structure and attributes of the graph.

\subsection{Known Perturbation Ratios}
In scenarios where the ratio of adversarial perturbations is known, a stratified approach is adopted to construct fine-tuned subgraphs for distinct types of attacks:
\subsubsection{Node Injection Attack: Subgraph Construction via Beta Wavelet Graph Neural Network (BWGNN)}
BWGNN\cite{tang2022rethinking} is employed to detect anomalies by identifying the spectral shift from low to high frequencies, indicative of node injections. The Beta distribution is used as a kernel function to create the Beta wavelet transform $W_{p, q}$, which serves as a band-pass filter with good spectral and spatial locality. The formulation of BWGNN can be summarized as:
\begin{equation}
	\begin{array}{l}
		W_{p, q}=\frac{1}{2 B(p+1, q+1)}\left(\frac{L}{2}\right)^{p}\left(I-\frac{L}{2}\right)^{q},
	\end{array}
	\label{BWGNN1}
\end{equation}
where $p, q \in \mathbb{N}^{+}$ control the spectral characteristics of the Beta wavelets, $B(p+1, q+1)$ is the Beta function, and $L$ is the normalized graph Laplacian matrix. Through Beta waveform transformation, the features of each node v are transformed to the frequency domain which helps in identifying the anomalous nodes. The transformed features can be expressed as $z_v = W_{p, q}(v)$. Specifically, the BWGNN may use Beta waveform transforms at different scales to capture signals at different frequencies. This means that each node v will have a set of feature vectors:
\begin{equation}
	\begin{array}{l}
		\mathbf{z}_{v}=\left[W_{0, C}\left(\mathbf{x}_{v}\right), W_{1, C-1}\left(\mathbf{x}_{v}\right), \ldots, W_{C, 0}\left(\mathbf{x}_{v}\right)\right],
	\end{array}
	\label{BWGNN2}
\end{equation}
where $C$ is the order of the transformation. The aggregated features ${z}_{v}$ are fed into an MLP $s(v)=\phi\left(\mathbf{z}_{v}\right)$ that learns a mapping from aggregated features to anomaly scores. The anomaly scores $s(v)$ are converted to anomaly probabilities ${\cal P}(v)$ by a sigmoid function:
\begin{equation}
	\begin{array}{l}
		p(v)=\sigma(s(v))=\frac{1}{1+e^{-s(v)}}.
	\end{array}
	\label{BWGNN3}
\end{equation}

Given $r$ be the threshold for anomaly detection. If $v \in {\cal V}$ with ${\cal P}(v) > r$ are considered anomalies, where ${\cal P}(v)$ is the anomaly probability assigned by BWGNN. Given a perturbation ratio $\zeta$, the number of nodes selected ${n'}=|{\cal V}| \times {\zeta}$. The higher ${\cal P}(v)$ is, the earlier it is selected. Subgraphs ${\cal G}_v$ are constructed by selecting two-hop neighbors ${N_2}(v)$ around each anomalous node $v$ in the anomalous nodes set ${\cal V}'$, excluding $v$ itself:
\begin{equation}
	\begin{array}{l}
		{{\cal G}_{v}} = {{N_2}(v)} \backslash \{v\},
	\end{array}
	\label{BWGNN4}
\end{equation}
Fine-tuned subgraph ${\cal G}_s$ is the union of all such ${{\cal G}_v}$:
\begin{equation}
	\begin{array}{l}
		{\cal G}_{s}=\bigcup_{v \in {\cal V}'} {{\cal G}_v},
	\end{array}
	\label{BWGNN5}
\end{equation}

\renewcommand{\algorithmicrequire}{\textbf{Input:}}
\renewcommand{\algorithmicensure}{\textbf{Output:}}
\begin{algorithm}[t]
	\caption{Construction of Fine-Tuned Subgraphs}
	\label{alg:construction}
	\begin{algorithmic}[1]
		\REQUIRE
		Poisoned graph ${\hat{\cal G}} = ({\hat V}, {\hat{\cal E}})$;
		perturbation ratios $\zeta, \vartheta, \kappa$.
		
		\ENSURE
		Fine-tuned subgraph ${\cal G}_{s}$.
		
		\STATE ${\cal G}_{s} \gets \emptyset$
		
		\IF{$Type = \text{node\_injection}$}
		\STATE Compute ${\cal P}(v)$ according to Equation \eqref{BWGNN3}
		\STATE ${\cal V}' \gets \{v \in {\hat V} \mid {\cal P}(v) > r \text{ and } |{\cal V}'| \leq \zeta |{\hat V}|\}$
		\STATE ${\cal G}_{s} \gets \bigcup_{v \in {\cal V}'} {N_2}(v) \setminus \{v\}$
		\ENDIF
		
		\IF{$Type = \text{feature\_modification}$}
		\STATE Compute ${\cal J}(v, u)$ according to Equation \eqref{jaccard1}
		\STATE ${\cal V}' \gets \{v \in {\hat V} \mid k_{v} > p \times |N_{1}(v)| \text{ and } |{\cal V}'| \leq \vartheta |{\hat V}|\}$
		\STATE Replace features of $v \in {\cal V}'$ with average of ${N_1}(v)$
		\STATE ${\cal G}_{s} \gets \bigcup_{v \in {\cal V}'} {\cal G}'_v$
		\ENDIF
		
		\IF{$Type = \text{edge\_modification}$}
		\STATE Compute $SimR(v,u)$ according to Equation \eqref{SimRank1}
		\STATE ${\cal E}' \gets \{e = (v,u) \in {\hat{\cal E}} \mid SimR(v,u) < \tau \text{ and } |{\cal E}'| \leq \kappa |{\hat{\cal E}}|\}$
		\STATE ${\cal G}_{s} \gets \bigcup_{e \in {\cal E}'} ({N_2}(v) \cup {N_2}(u)) \setminus \{e\}$
		\ENDIF
		
		\RETURN ${\cal G}_{s}$
	\end{algorithmic}
\end{algorithm}

\subsubsection{Feature Modification Attack: Subgraph Construction via Jaccard Similarity}
The Jaccard similarity coefficient ${\cal J}(v,u)$ measures the dissimilarity between a node's features $F(v)$ and those of its neighbors $F(u)$. For each node $v$, the jaccard similarity with its one-hop neighbors ${N_1}(v)$ is calculated as:
\begin{equation}
	\begin{array}{l}
		{\cal J}(v, u)=\frac{|F(v) \cap F(u)|}{|F(v) \cup F(u)|},
	\end{array}
	\label{jaccard1}
\end{equation}
where $|F(v) \cap F(u)|$ denotes the size of the intersection of the feature sets of node $v$ and neighbor $u$. If the Jaccard Similarity ${\cal J}(v, u)$ with any neigbor $u$ is below the threshold $r$, the condition for node $v$ being considered anomalous is:
\begin{equation}
	\begin{array}{l}
		k_{v}=\sum_{u \in N_{1}(v)} 1_{J(v, u)<r},
	\end{array}
	\label{jaccard2}
\end{equation}
where $1_{J(v, n)<r}$ is an indicator function that equals 1 if ${\cal J}(v, u) < r$ and 0 otherwise. If $k_{v}$ exceeds a certain proportion $p$ of the node's one-hop neighbors(e.g., more than 50\% of neighbors are dissimilar):
\begin{equation}
	\begin{array}{l}
		k_{v}>p \times\left|N_{1}(v)\right|,
	\end{array}
	\label{jaccard3}
\end{equation}
mark node $v$ as anomalous. Where the perturbation ratio is known, the detected anomalous nodes also need to be selected by the perturbation ratios $\vartheta$. The higher $k_{v}$ is, the earlier it is selected. For each node $v$ in the anomalous nodes set ${\cal V}'$, construct a subgraph ${\cal G}_v$ that includes the two-hop neigbors ${N_2}(v)$. Since direct removal of node features modified nodes breaks the structural features of the original graph, replace the features of $v$ with the average features of its one-hop neighbors ${N_1}(v)$ to form ${{\cal G}'}_v$. Fine-tuned subgraph ${\cal G}_s$ is the union of all such ${{\cal G}'}_v$:
\begin{equation}
	\begin{array}{l}
		{{\cal G}_{s}} = {\bigcup}_{v \in {{\cal V}'}} {{\cal G}'}_v,
	\end{array}
	\label{jaccard4}
\end{equation}

\begin{figure*}
	\centering
	\includegraphics[width=1.0\textwidth]{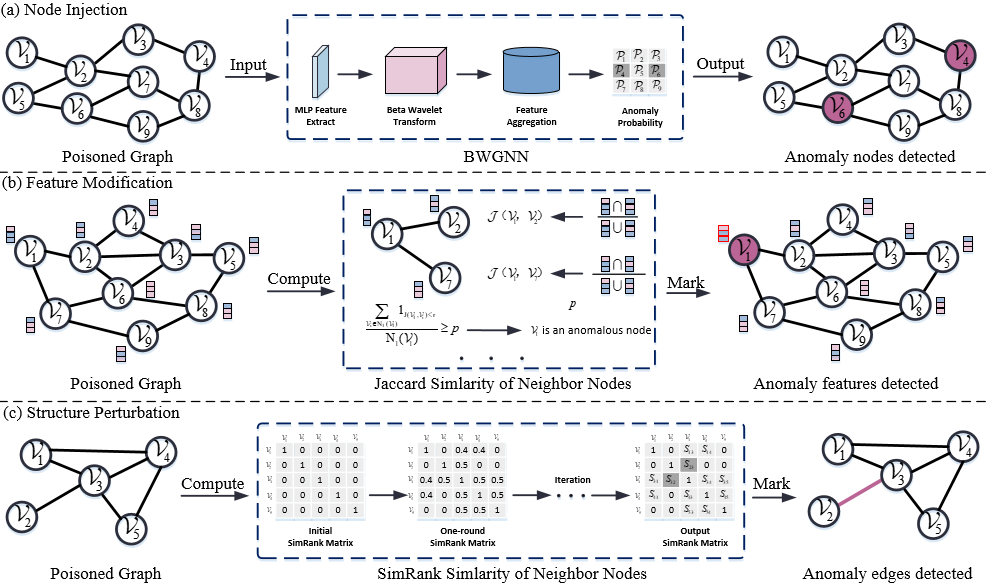}
	\caption{Anomaly detection methods used in this paper. (a)For node injection attacks, we use BWGNN to detect anomalous injected nodes. (b)For feature modification attacks, we identify the anomalous nodes by calculating the jaccard similarity between each node and its neighboring nodes. (c)For structure perturbation attacks, we compute the simrank similarity between each node and its neighboring nodes as a way to identify maliciously edges.}
	\label{subgraph extraction}
\end{figure*}

\subsubsection{Structure Perturbation Attack: Subgraph Construction via SimRank}
In addressing structure perturbation attacks, we employ the SimRank theory to quantify the similarity between pairs of nodes, which is integral to identifying anomalous edges indicative of adversarial modifications. The SimRank similarity $SimR(v,u)$ between any two node $v$ and $u$ in graph $\cal G$ is defined by the recursive relation:
\begin{equation}
	\begin{array}{l}
		SimR(v,u) = \frac{1}{2} \sum_{x \in {N_1}(v)} \sum_{y \in {N_1}(u)} \frac{SimR(x,y)}{|{N_1}(v)| \times|{N_1}(u)|},
	\end{array}
	\label{SimRank1}
\end{equation}
where ${N_1}(v)$ and ${N_1}(u)$ denote the one-hop neighbor sets of nodes $v$ and $u$, respectively, and $SimR(x,y)$ represents the similarity between nodes $x$ and $y$. The inital conditions are set such that $SimR(v,v) = 1$ for all nodes $v$, and $SimR(v,u) = 0$ for all distinct nodes $v$ and $u$. The iterative computation of $SimR(v,u)$ converages to stable state, reflecting the inherent structural similarity between nodes. An edge $e = (v,u)$ is deemed anomalous if its associated SimRank value falls below a predetermined threshold $\tau$, suggesting a low probability of the edge existing in the graph's legitimate structure.

Given the set of anomalous edges ${\cal E}'$, in case of knowing perturbation ratios, the set of anomalous edges $E''$ is further selected based on knowing perturbation ratios. The higer $SimR(v,u)$ is, the earlier it is selected. For each anomalous edge $e \in {{\cal E}''}$, the subgraph $G_e$ is formed by aggregating the two-hop neighbors of the nodes connected by $e$ and excluding the anomalous edge $e$ from the original graph:
\begin{equation}
	\begin{array}{l}
		{{\cal G}_{e}} = {({N_2}(v) \cup {N_2}(u))} \backslash \{e\},
	\end{array}
	\label{SimRank2}
\end{equation}
where ${N_2}(v)$ and ${N_2}(v)$ represent the two-hop neighborhoods of nodes $v$ and $u$, respectively. The fine-tuned subgraph ${\cal G}_s$ is the union of all such subgraphs ${\cal G}_e$ associated with the detected anomalies:
\begin{equation}
	\begin{array}{l}
		{{\cal G}_{s}} = {\bigcup}_{e \in {{\cal E}''}} {{\cal G}}_e,
	\end{array}
	\label{SimRank3}
\end{equation}

\subsection{Complete Knowledge of Perturbations}
When complete information about the adversarial perturbations is available, the detection process is bypassed, and subgraphs are directly constructed around the identified perturbations. For node injection, subgraphs are centered on the injected nodes $v_i$, excluding them and including their two-hop neighbors ${N_2}(v_i)$, forming ${\cal G}_{s} = \bigcup_{i} N_{2}(v_i) \backslash\{v_i\}$. For feature modification, subgraphs are formed around nodes $v_j$ with modified features, adjusting their features to the average of one-hop neigbors ${N_1}(v_j)$, forming ${{\cal G}_{s}} = \bigcup_{j}\left(N_{2}\left(v_{j}\right)\right)$. For edge addition, subgraph are created around the added edges $e_k$,  excluding these edges and including the two-hop neighbors of the connected nodes $u_k$ and $v_k$, forming ${\cal G}_{s} = \bigcup_{k}({N_2}(u_k) \cup {N_2}(v_k)) \backslash \{e_k\}$.

\subsection{No Knowledge of Perturbations}
In the absence of specific perturbations knowledge, anomaly detection parallels the known perturbation ratio scenario, except that there is no further filtering step. BWGNN, Jaccard Similarity, and SimRank algorithms are still utilized to identify anomalous nodes and edges. Specifically, for node injection attacks and feature modification attacks, attack detection methods detects poisoned nodes without the need for ratio-based selection. All nodes identified as anomalous are used to construct subgraphs ${\cal G}_v$ centered around each node and its two-hop neighbors. For structure perturbation, anomalous edges identified by attack detection methods also have no need for ratio-based filtering. All identified  anomalous edges are used to construct subgraphs ${\cal G}_v$ centered around each edge and two-hop neighbors of the two nodes it connects.

\begin{algorithm}[t]
	\caption{Parameters Optimization}
	\label{alg:training}
	\begin{algorithmic}[1]
		\REQUIRE: Fine-tuned subgraph ${\cal G}_{s}$, parameters of poisoned GCN $\theta$
		\ENSURE: Optimized parameters $\hat {\theta}$;
		\STATE initialize GCN model ${\cal M}(\cdot)$ with parameters $\theta$;
		\STATE Compute ${{\cal H}^{(1)}}$ and ${{\cal H}^{(2)}}$ with the subgraph ${\cal G}_{s}$;
		\STATE Calculate the loss $\mathcal{L}$ with the true labels $\cal Y$ and predictions ${{\cal H}^{(2)}}$;
		\STATE Compute the gradient $\nabla_{\theta} \mathcal{L}$ of the loss with respect to $\theta$;
		\STATE Update the model parameters: ${\hat {\theta}}= \theta -\eta \cdot \nabla_{\theta} \mathcal{L}$;
		\REPEAT
		\STATE Update ${{\cal H}^{(1)}}$, ${{\cal H}^{(2)}}$, and $\nabla_{\theta} \mathcal{L}$ with the new $\theta$
		\STATE Update ${\hat {\theta}}$ with the learning rate $\eta$;
		\UNTIL Convergence or maximum iterations reached
		\RETURN{Optimized parameters $\hat {\theta}$}
	\end{algorithmic}
\end{algorithm}

\subsection{Fine-tuning the Poisoned GNN}
We employ fine-tuned subgraph ${\cal G}_s = ({\cal A}_s,{\cal X}_s)$ as the input for model repair, optimizing the model's parameters. The adjacency and feature matrices are denoted by ${\cal A}_s$ and ${\cal X}_s$, respectively, with $\cal Y$ representing the true node labels. The model parameters are updated without prior adjustment, as the subgraph is based on the pre-trained model's dataset. For a two-layer GCN, the feature learning process is as follows:
\begin{equation}
	\begin{array}{l}
		{{\cal H}^{(1)}} = \sigma \left( {{{\tilde {\cal D}}^{ - \frac{1}{2}}}{{\cal A}_s}{{\tilde {\cal D}}^{ - \frac{1}{2}}}{{\cal X}_s}{{\cal W}^{(0)}}} \right),\\
		{{\cal H}^{(2)}} = \sigma \left( {{{\tilde {\cal D}}^{ - \frac{1}{2}}}{{\cal A}_s}{{\tilde {\cal D}}^{ - \frac{1}{2}}}{{\cal H}^{(1)}}{{\cal W}^{(1)}}} \right),
	\end{array}
	\label{fine-tuning1}
\end{equation}
where ${\cal W}^{(0)}$ and ${\cal W}^{(1)}$ are pre-trained weights. The loss function measures the discrepancy between the model's output and true labels:
\begin{equation}
	\begin{array}{l}
		\mathcal{L}(\theta, y, h)=-\sum_{i} y_{i} \log \left(h_{i}\right)+\left(1-y_{i}\right) \log \left(1-h_{i}\right),
	\end{array}
	\label{fine-tuning2}
\end{equation}
with $h_i$ from $H_{(2)}=\{h_1,...,h_n\}$, $y_i$ from ${\cal Y}=\{{y_1},...,{y_n}\}$, and $\theta$ comprising $W^{(0)}$ and $W^{(1)}$. Parameter update $\theta$ is achieved via gradient computation:
\begin{equation}
	\begin{array}{l}
		\nabla_{\theta} \mathcal{L}=\frac{\partial \mathcal{L}}{\partial h} \frac{\partial h}{\partial \theta},
	\end{array}
	\label{fine-tuning3}
\end{equation}
followed by optimization using the learning rate $\eta$:
\begin{equation}
	\begin{array}{l}
		{\hat {\theta}}= \theta -\eta \cdot \nabla_{\theta} \mathcal{L},
	\end{array}
	\label{fine-tuning4}
\end{equation}
The entire algorithm for optimize the model parameters is summarized as Algorithm 2. The whole process of fine-tuning is actually equivalent to the poisoned GNN training again to ``correct'' the parameters of the model.

\section{Unlearning Validation}
The objective of our unlearning method is to diminish the influence of poisoned nodes on its neighboring nodes and the overall graph. To validate this, we propose a validating ideas that links the changes in probability predictions of neighboring nodes to the unlearning of the target node.

\begin{algorithm}[t]
	\caption{Unlearning Validation Algorithm}
	\label{alg:unlearning_validation}
	
	\begin{algorithmic}[1]
		\REQUIRE Original graph $\mathcal{G}$, fine-tuned subgraph $\tilde{\mathcal{G}}$, poisoned node $v_i$, neighbor node $v_j$, and classes $k_1, k_2$.
		\ENSURE Validation of unlearning effectiveness for node $v_i$.
		
		\STATE Initialize the output of the $(l+1)$th GNN layer for $\mathcal{G}$ as ${H}^{(l+1)}$.
		\STATE Initialize softmax probability distributions $\mathcal{O}_{v_i}$ and $\mathcal{O}_{v_j}$ for nodes $v_i$ and $v_j$ respectively.
		\FORALL{classes $k \in \{k_1, k_2\}$}
		\STATE Calculate $\mathcal{O}_{v_{i},k}$ and $\mathcal{O}_{v_{j},k}$ based on the GNN output.
		\ENDFOR
		\STATE Predict classes $\hat{y}_{v_i}$ and $\hat{y}_{v_j}$ as the indices of the maximum values in $\mathcal{O}_{v_i}$ and $\mathcal{O}_{v_j}$ respectively.
		\STATE Compute the corresponding softmax distributions $\tilde{\mathcal{O}}_{v_i}$ and $\tilde{\mathcal{O}}_{v_j}$ for the fine-tuned subgraph $\tilde{\mathcal{G}}$.
		\FORALL{classes $k \in \{k_1, k_2\}$}
		\STATE Compute the probability differences $\Delta\mathcal{O}_{v_{j},k} = \mathcal{O}_{v_{j},k} - \tilde{\mathcal{O}}_{v_{j},k}$.
		\ENDFOR
		\IF{$\hat{y}_{v_i} = k_1$ and $\hat{y}_{v_j} = k_1$ and $\Delta\mathcal{O}_{v_{j},k_1} > 0$}
		\RETURN "Unlearning effective for $v_i$."
		\ELSIF{$\hat{y}_{v_i} = k_1$ and $\hat{y}_{v_j} = k_2$ and $\Delta\mathcal{O}_{v_{j},k_2} \leq 0$}
		\RETURN "Unlearning effective for $v_i$."
		\ENDIF
		\RETURN "Unlearning not effective for $v_i$."
	\end{algorithmic}
\end{algorithm}
For GNN, if the output of the last layer is ${\cal H}^{(l+1)}$, then the probability distribution obtained by the softmax function can be formulated as:
\begin{equation}
	\label{val1}
	{\cal O}=Softmax({\cal H}^{(l+1)}),
\end{equation}
For one node $v_{i}$ and class $k$, the output $\cal O$ can be formulated as:
\begin{equation}
	\label{val2}
	{\cal O}_{v_{i},k}=Softmax({h}^{(l+1)}),
\end{equation}
Normally, for node $v_{i}$, we choose the class k with the highest probability output as the prediction class:
\begin{equation}
	\label{val3}
	{\hat y_{v_{i}}} = \arg {\max _k}{{\cal O}_{v_{i},k}}.
\end{equation}

After unlearning, the influence of a poisoned node $v_i$ and its associated poisoned edges on its neighboring nodes $v_j$ should be diminished. The reduction is expected to manifest as changes in the probability predictions ${\cal O}_{v_{j},k}$ for $v_j$ before and after unlearning.

The validation criteria are based on the comparative analysis of the probability distributions ${\cal O}_{v_{j},k}$ and ${\tilde{\cal O}}_{v_{j},k}$ where ${\tilde{\cal O}}_{v_{j},k}$ represents the distribution post-unlearning.

\textbf{Influence Reduction for Same-Class Neighbors}: if a neighbor $v_j$ shares the same class $k_1$ as the poisoned node $v_i$, the unlearning should result in a decreased likelihood of $v_j$ being classified as $k_1$:
\begin{equation}
	\label{val4}
	{{{\cal O}_{{v_j},{k_1}}} > {{\tilde {\cal O}}_{{v_j},{k_1}}},{\;}y_{{v_j}} = y_{{v_i}} = {k_1}}.
\end{equation}

\textbf{Influence Reduction for Different-Class Neighbors}: if $v_j$ belongs to a different class $k_2$ than the poisoned node ${v_i}(class k_1)$, the unlearning should not alter the likelihood of $v_j$ being classified as $k_2$, but should reduce the incorrect classification likelihood as $k_1$:
\begin{equation}
	\label{val5}
	{{{\cal O}_{{v_j},{k_2}}} \le {{\tilde {\cal O}}_{{v_j},{k_2}}} {\;}{\text or}{\;} {{\cal O}_{{v_j},{k_1}}} > {{\tilde {\cal O}}_{{v_j},{k_1}}},{\;}y_{{v_i}} = {k_1},{\;}y_{{v_j}} = {k_2}}.
\end{equation}

These criteria ensure that the unlearning of the poisoned node $v_i$ and its edges has effectively reduced their influence on the neighboring nodes, as evidenced by the changes in the probability prediction for $v_j$.

\section{Model Comparison and Analysis}
\subsection{Comparison Discussion}
\subsubsection{Connection to SAFE}
SAFE\cite{dukler2023safe} introduces a shard graph to enable selective forgetting in large models. By partitioning data into shards and training independent models, it minimizes the cost of removing specific training samples' influence. SAFE leverages synergistic information between shards to enhance accuracy without significantly increasing the forgetting cost. It employs lightweight adapters for efficient training and reuse of computations, allowing it to manage more shards and reduce forgetting costs compared to existing methods. In contrast, our proposed GraphMU presents a specialized repair framework for GNNs post-adversarial attack. This targeted approach to model repair and robustness enhancement is a significant advantage for GNN security in adversarial contexts.

\subsubsection{Connection to GNNDELETE}
GNNDELETE\cite{cheng2022gnndelete} presents a model-agnostic approach to unlearning in GNNs by focusing on the properties of ``Deleted Edge Consistency" and ``Neighborhood Influence", allowing for efficient and targeted data removal without full model retraining. This method stands out for its ability to handle various deletion tasks and maintain the integrity of the remaining graph structure. In comparison, our proposed GraphMU method offers a specialized strategy for repairing GNNs that have been specifically compromised by adversarial attacks. The advantage of our work lies in its more use of information from graph used during training to guide the fine-tuning process, ensuring that the GNN not only forgets the adversarial samples but also retains its robustness against such attacks.

\subsection{Time Complexity Analysis}
The time cost of GraphMU mainly originates in constructing fine-tuned subgraph. The overall time complexity of the algorithm will depend on which attack type is being executed, as each has its own set of computationally intensive steps. However, if we consider the most computationally expensive parts of each attack type, we can estimate an upper bound for the overall time complexity.

\textbf{Node Injection Attack}:the complexity primarily arises from the computation of the BWGNN features of each node, which involves an MLP with a complexity of $O(n \times D \times L)$, where $n$ is the number of nodes, $D$ is the feature vector size, and $L$ represents the complexity of the MLP layers. Additionally, selecting the anomalous nodes involves sorting or priority queue operations, adding a complexity of $O(n \times log(n))$. Constructing the subgraph for each anomalous node has a complexity of $O(n \times d)$, where $d$ is the average degree of the nodes. Therefore, the dominant term for this attack type is $O(n \times D \times L)$.

\textbf{Feature Modification Attack}: The computation of Jaccard similarities for each node and its neighbors is the most expensive step, $O(n \times d^2)$. The selection of anomalous nodes and the construction of subgraphs are less siginificant in comparison. Thus, the dominant term for this attack type is $O(n \times d^2)$.

\textbf{Structure Perturbation Attack}: The iterative computation of SimRank similarities between all pairs of nodes is the most intensive, $O(m \times n^2)$. The selection of anomalous edges and the construction of subgraphs follow, with complexities $O(m \times n \times log(n))$ and $O(m \times n \times d)$, respectively. Here, the dominant term is $O(m \times n^2)$.

\section{Experiments}
\subsection{Experimental Setup}
\paragraph{Datasets}
To evaluate the effectiveness of our method in repairing poisoned GCN, we conduct extensive experiment on four representative benchmark datasets, including Cora, Citeseer,  Pubmed\cite{mccallum2000automating}, and Cora-ML\cite{sen2008collective}. These datasets are all citation graphs, in which each node represents scientific literature, and edges between nodes represent the citation relationship of scientific literature. The statistics of the datasets are shown in Table \ref{tab:datasets}.
\begin{table}[!h]
	\caption{Dataset statistics\label{tab:datasets}}
	\centering
	\begin{tabular}{c | c c c c}
		\hline
		Dataset & Nodes & Edges & Features & Labels \\
		\hline
		Cora & 2708 & 5429 & 1433 & 7 \\
		Citeseer & 3312 & 4732 & 3703 & 6 \\
		Cora-ML & 2810 & 7981 & 2879 & 7 \\
		Pubmed & 19717 & 44324 & 500 & 3 \\
		\hline
	\end{tabular}
\end{table}

\begin{figure*}[htbp]
	\centering
	\subfigure[]{
		\includegraphics[width=1.0\textwidth]{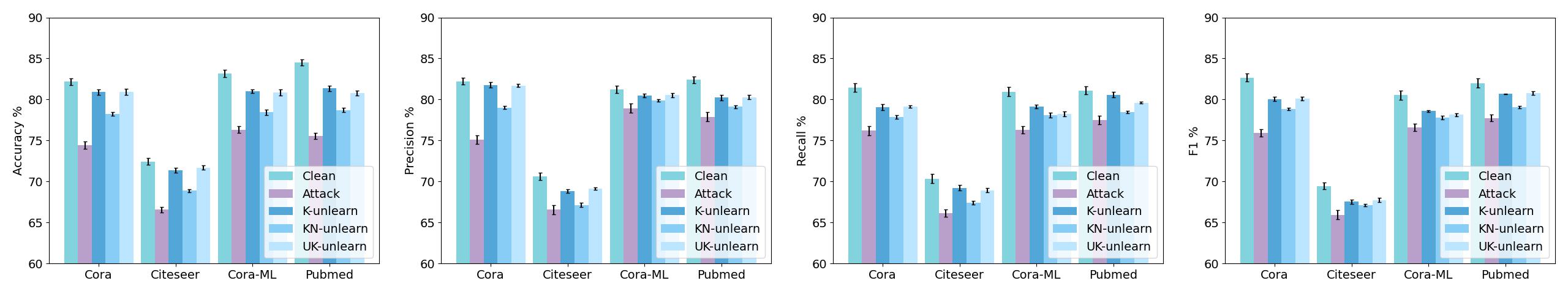}
	}
	\subfigure[]{
		\includegraphics[width=1.0\textwidth]{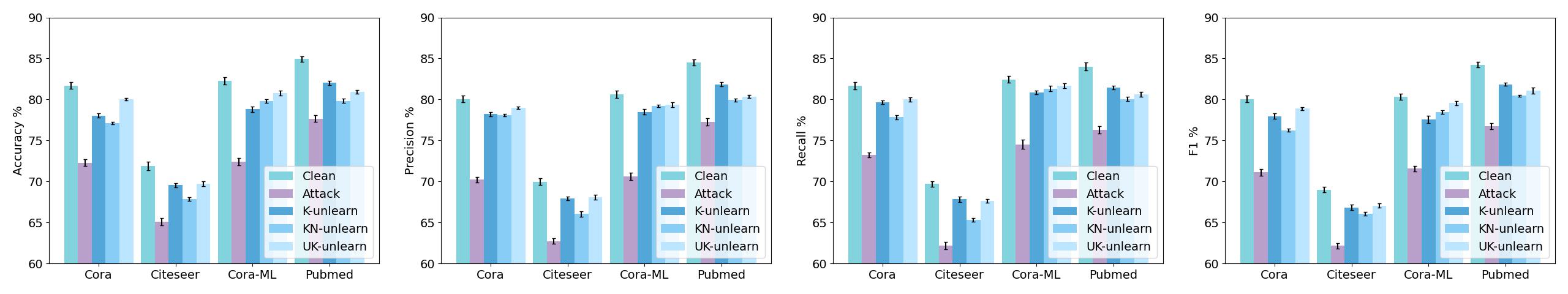}
	}
	\subfigure[]{
		\includegraphics[width=1.0\textwidth]{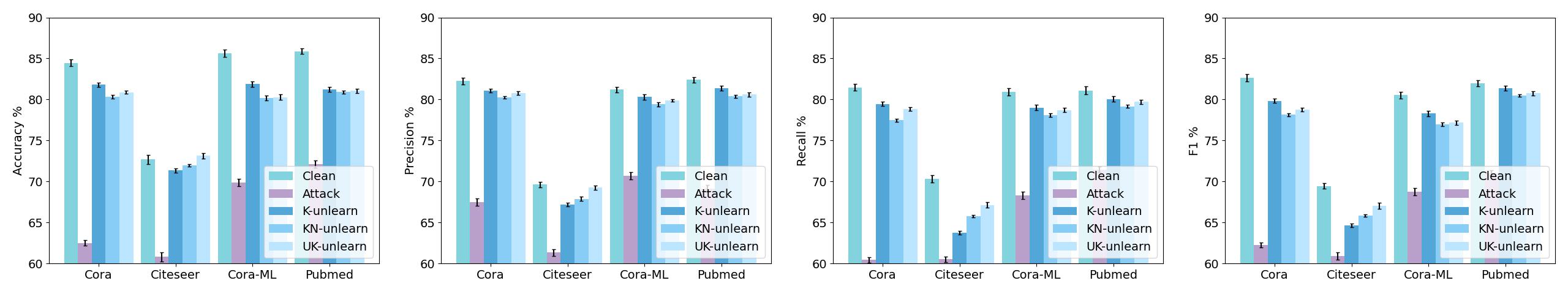}
	}
	\subfigure[]{
		\includegraphics[width=1.0\textwidth]{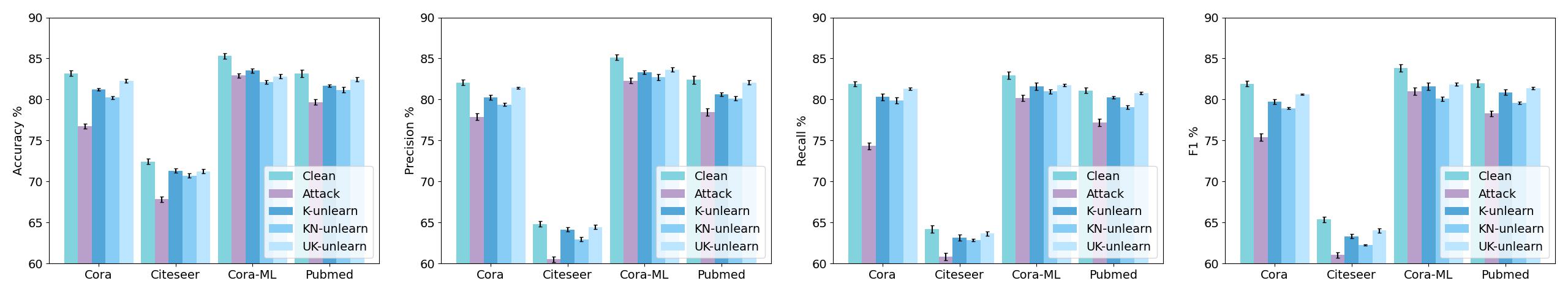}
	}
	\caption{The effectiveness of GraphMU in repairing the poisoned GCN under conditions of 2-hop fine-tuned subgraph and 5-round fine-tuning. (a) The effectiveness of GraphMU in repairing the poisoned GCN under Nettack. (b) The effectiveness of GraphMU in repairing the poisoned GCN under GANI. (c) The effectiveness of GraphMU in repairing the poisoned GCN under SGA. (d) The effectiveness of GraphMU in repairing the poisoned GCN under Min-Max.}
	\label{unlearning}
\end{figure*}

\begin{figure*}[htbp]
	\centering
	\subfigure[]{
		\includegraphics[width=0.23\textwidth]{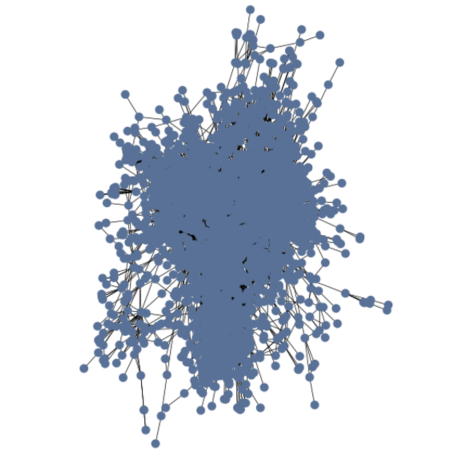}
	}
	\subfigure[]{
		\includegraphics[width=0.23\textwidth]{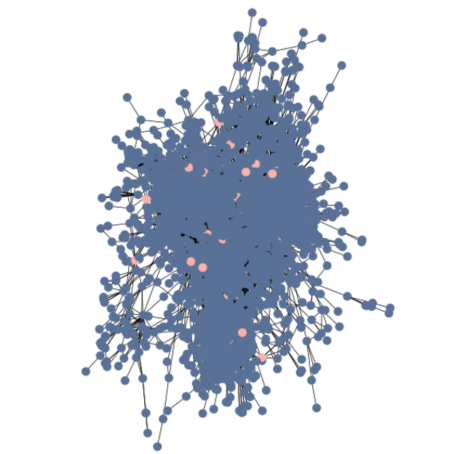}
	}
	\subfigure[]{
		\includegraphics[width=0.23\textwidth]{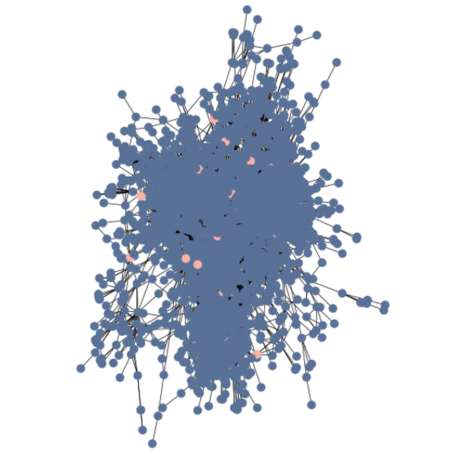}
	}
	\subfigure[]{
		\includegraphics[width=0.23\textwidth]{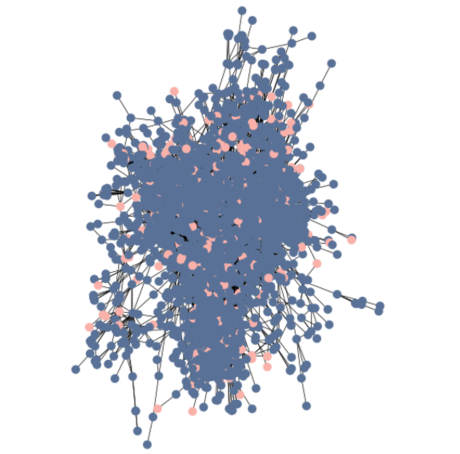}
	}
	\caption{Visualization of Cora-ML distribution under Nettack. (a) Clean graph. (b) Graph including poisoned samples. (c) Graph including poisoned samples detected by anomaly detection under the limitations of number. (d) Graph including poisoned samples detected by anomaly detection}
	\label{Visualization}
\end{figure*}

\begin{table*}[htbp]
	\caption{Computational time costs (s) of GraphMU on four datasets under four adversarial attack methods. The parameters for setting up this experiment: 2-hop fine-tuned subgraph, 5 rounds of fine-tuning. The scenario for setting up this experiment: known all information with regard to perturbations}
	\begin{tabular}{p{1.2cm}p{0.8cm}|p{0.8cm}p{0.7cm}p{0.6cm}p{1.25cm}|p{0.8cm}p{0.7cm}p{0.6cm}p{1.25cm}|p{0.8cm}p{0.7cm}p{0.6cm}p{1.25cm}}
		\hline
		\multicolumn{2}{c|}{}                   & \multicolumn{4}{c|}{Known all perturbations} & \multicolumn{4}{c|}{Known perturbation ratios} & \multicolumn{4}{c}{Unknown any perturbation information} \\ \cline{3-14}
		Dataset                       & Retrain & Nettack     & GANI     & SGA      & Min-Max    & Nettack      & GANI       & SGA       & Min-Max     & Nettack       & GANI       & SGA        & Min-Max      \\ \hline
		\multicolumn{1}{c|}{Cora}     & 6.33    & 0.66        & 0.42     & 2.46     & 2.66       & 0.61         & 0.41       & 2.33      & 2.74        & 0.84          & 0.72       & 3.69       & 4.44         \\
		\multicolumn{1}{c|}{Cora-ML}  & 7.26    & 1.27        & 0.37     & 3.27     & 3.12       & 1.22         & 0.33       & 3.21      & 3.28        & 1.76          & 0.88       & 4.77       & 5.62         \\
		\multicolumn{1}{c|}{Citeseer} & 7.52    & 0.62        & 0.41     & 2.61     & 2.41       & 0.68         & 0.46       & 2.47      & 2.49        & 0.87          & 0.69       & 3.53       & 4.18         \\
		\multicolumn{1}{c|}{Pubmed}   & 19.21   & 55.60       & 11.76    & 11.74    & 4.18       & 53.16        & 11.12      & 11.15     & 4.66        & 83.02         & 17.15      & 14.64      & 7.63         \\ \hline
	\end{tabular}
	\label{tab:running time}
\end{table*}
\paragraph{Attack setup}
To highlight the outstanding performance in repairing the poisoned GCN of our method, we attack GCN with various adversarial attacks methods on GNNs to make it poisoned, and then repair poisoned GCN with our method. More detailed descriptions of the graph adversarial attacks are follows:
\begin{itemize}
	\item Nettack\cite{zugner2018adversarial}: The algorithm employs a strategic incremental perturbation to circumvent local optima in the optimization of graph structures. In this paper, we use it as a paradigm for feature modification attacks.
	\item GANI\cite{fang2024gani}: This strategy improves the stealth and potency of GNN adversarial attacks via a sophisticated node injection technique and optimizes global attack performance with a genetic algorithm for adjacency selection, maintaining perturbation imperceptibility in structure and features. we make use of it as a paradigm for node injection attacks in our work.
	\item SGA\cite{li2021adversarial}: This approach enhances efficiency in adversarial attacks by targeting a smaller subgraph of k-hop neighbors around the target node within a multi-stage framework. It also incorporates a scale factor to mitigate gradient vanishing, boosting attack performance. In this paper, we utilize it as an example for structures perturbation attacks.
	\item Min-Max\cite{xu2019topology}: This method introduces a gradient-based attack generation framework from an optimization perspective, utilizing first-order gradient information to guide the perturbation of graph topology, effectively avoiding local optima. In our work, we use it as an example of mixing multiple attack scenarios, including node feature modification attacks and structures perturbation attacks.
\end{itemize}

\begin{figure*}[htbp]
	\centering
	\subfigure[]{
		\includegraphics[width=1.0\textwidth]{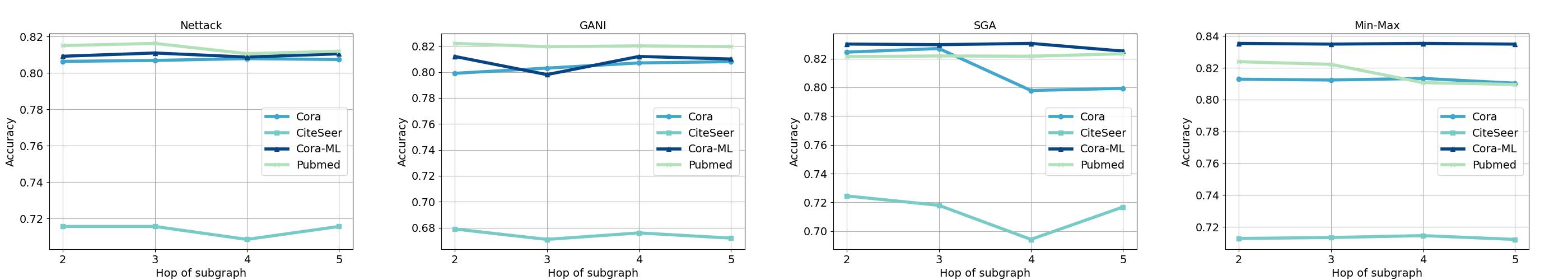}
	}
	\subfigure[]{
		\includegraphics[width=1.0\textwidth]{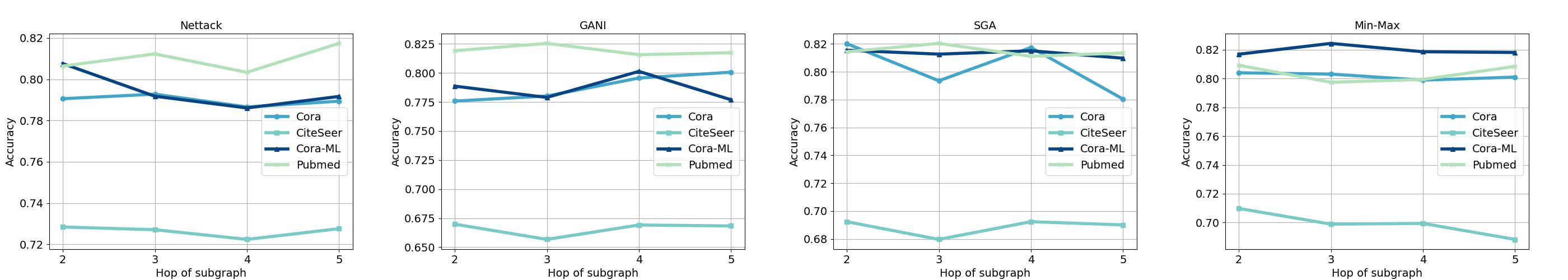}
	}
	\subfigure[]{
		\includegraphics[width=1.0\textwidth]{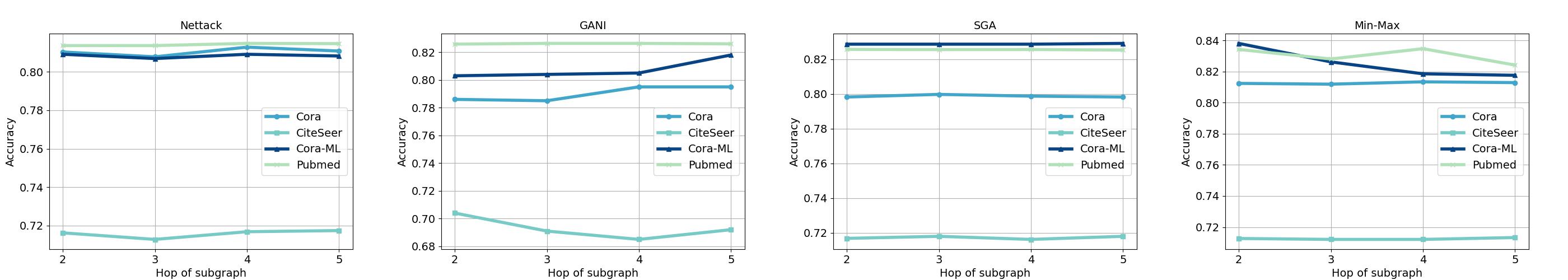}
	}
	\caption{The effect of the hop of fine-tuned subgraph construction on repairing performance under different scenarios. (a) poisoned samples are all known. (b) only the number of perturbations is known. (c) poisoned samples are completely unknown.}
	\label{hop-effect}
\end{figure*}

\begin{figure*}[htbp]
	\centering
	\subfigure[]{
		\includegraphics[width=1.0\textwidth]{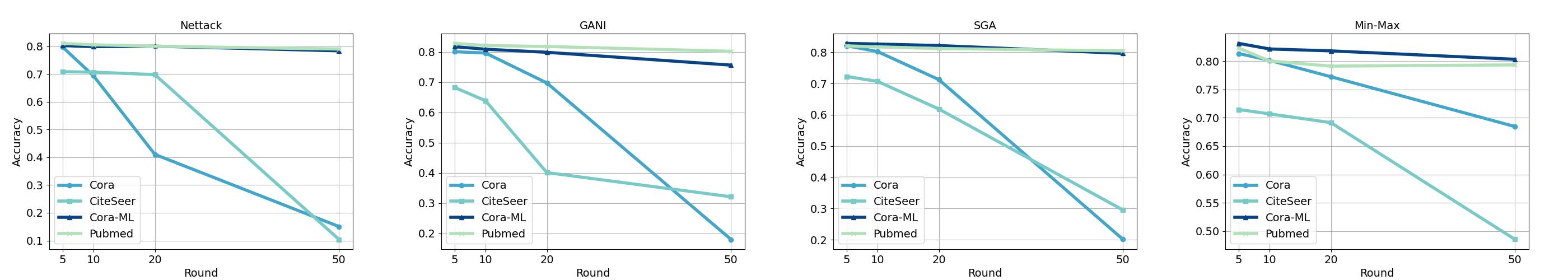}
	}
	\subfigure[]{
		\includegraphics[width=1.0\textwidth]{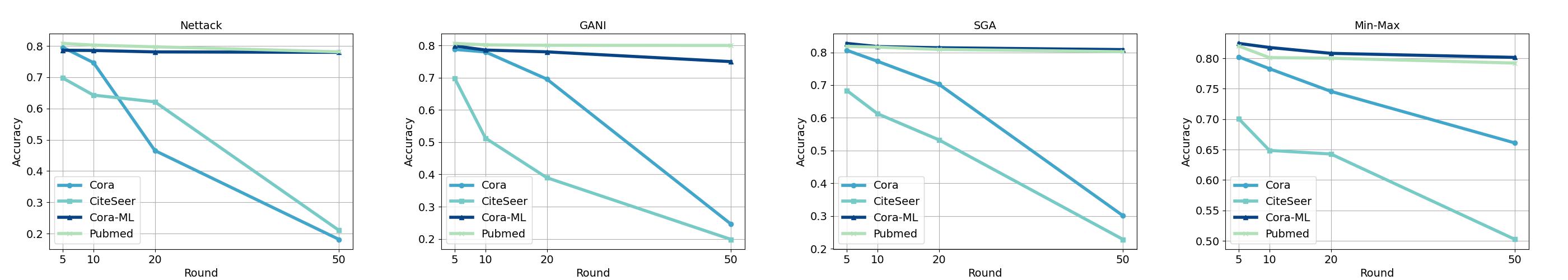}
	}
	\subfigure[]{
		\includegraphics[width=1.0\textwidth]{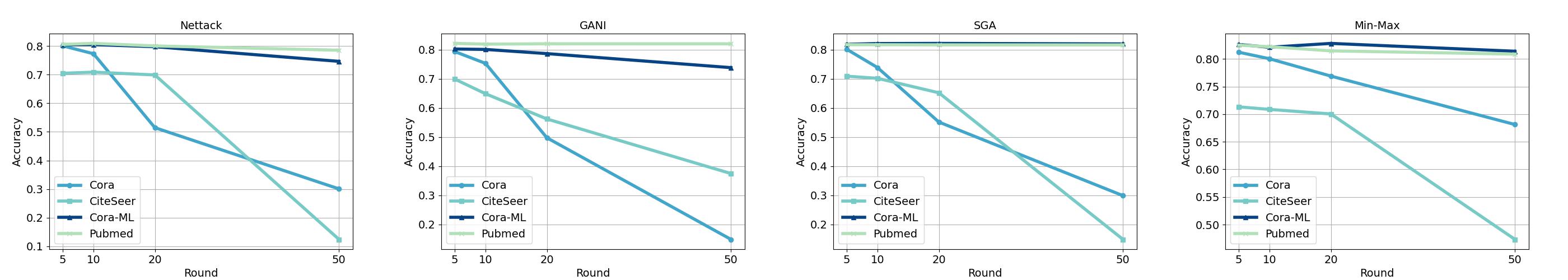}
	}
	\caption{The effect of the rounds of fine-tuning on repairing performance under different scenarios. (a) poisoned samples are all known. (b) only the number of perturbations is known. (c) poisoned samples are completely unknown.}
	\label{round-effect}
\end{figure*}
\paragraph{Target model setup}
Given that most adversarial attacks on GNNs target specific GNNs, we specifically targets the Graph Convolution Network(GCN) as the objective of all sorts of adversarial attacks.

\paragraph{Evaluation Metrics}
We evaluate the effectiveness of our method in repairing poisoned GCN using four metrics. The goal of improving the robustness of GNNs is to increase the values of these metrics.
\begin{itemize}
	\item Accuracy:
	\begin{equation}
		\label{Accuracy}
		Accuracy = \frac{{TP + TN}}{{TP + FN + FP + TN}}.
	\end{equation}
	\item Precision:
	\begin{equation}
		\label{Precision}
		Precision = \frac{{TP}}{{TP + FP}}.
	\end{equation}
	\item Recall:
	\begin{equation}
		\label{Recall}
		Recall = \frac{{TP}}{{TP + FN}}.
	\end{equation}
	\item F1 Score:
	\begin{equation}
		\label{F1Score}
		F1 Score = \frac{{2 \times Precision \times Recall}}{{Precision + Recall}}.
	\end{equation}
\end{itemize}

\subsection{Effectiveness of GraphMU}
In this section, the effectiveness of GraphMU in repairing poisoned GCN is tested. We make GCN poisoned with distinct adversarial attack methods respectively, and then use our method to repair it. Fig. \ref{unlearning} shows the performance of our proposed GraphMU on four datasets under four adversarial attack methods. Specifically, we set three scenarios based on the level of knowledge of perturbations: ({\romannumeral1}) K-unlearn: Known all information about perturbations, ({\romannumeral2})KN-unlearn: Known only the ratio of perturvations, and ({\romannumeral3})UK-unlearn: Unknown any information about perturbations. In the case of K-unlearn, we have complete information with respect to perturbations, thus we can directly construct the fine-tuned subgraph. For both the KN-unlearn and UK-unlearn cases, we utilize the Beta Wavelet Graph Neural Network (BWGNN) \cite{tang2022rethinking} for anomaly detection to identify the poisoned samples. For the KN-unlearn scenario, the number of detected anomalous samples is limited to the number of real poisoned samples, whereas there is no limit on the number of anomalous samples in the UK-unlearn scenario.

\begin{figure*}[htbp]
	\centering
	\subfigure[]{
		\includegraphics[width=1.0\textwidth]{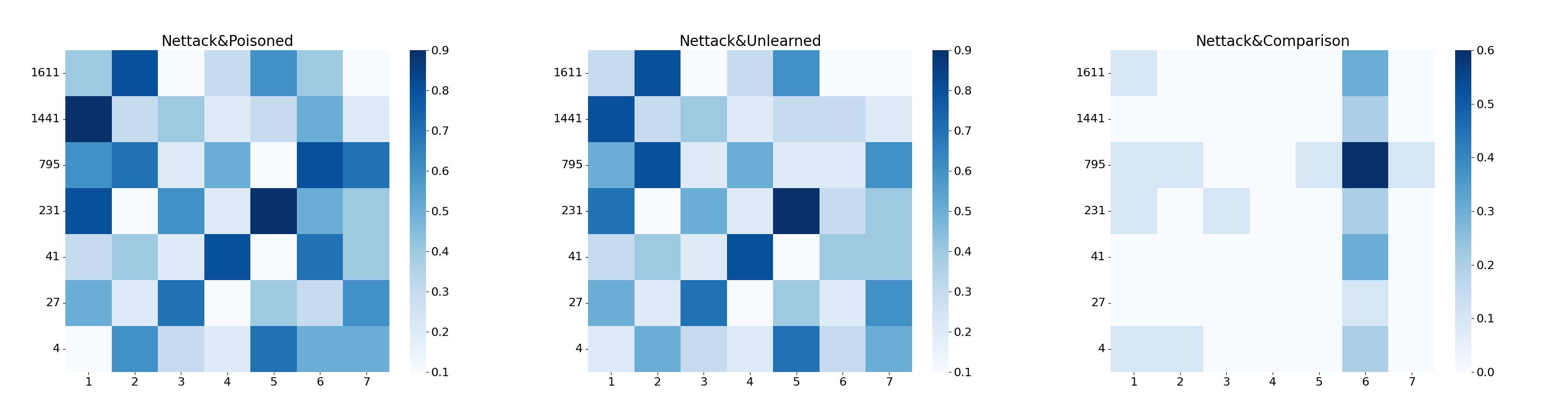}
	}
	\subfigure[]{
		\includegraphics[width=1.0\textwidth]{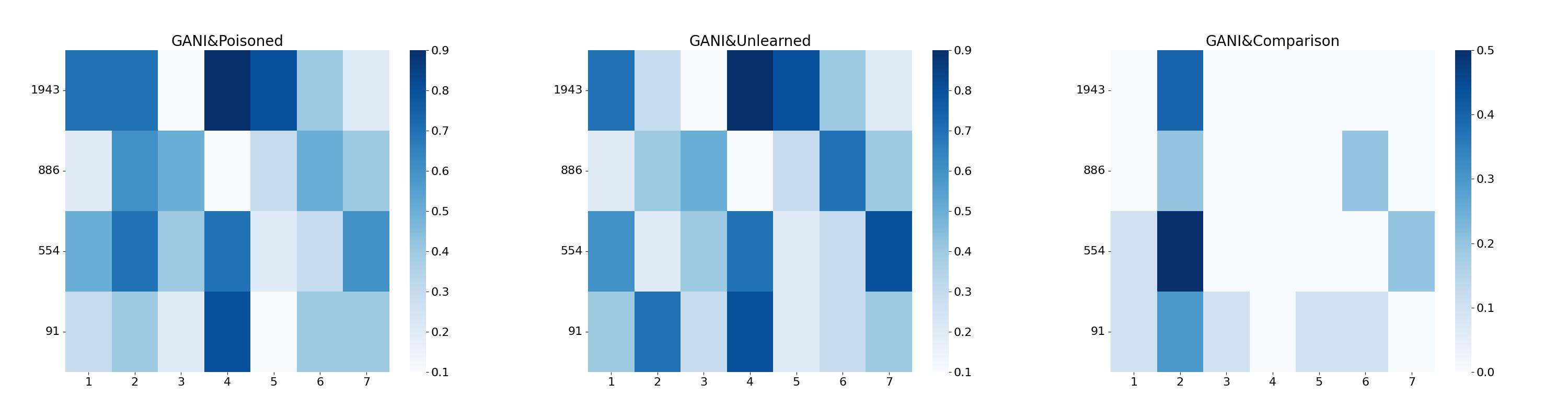}
	}
	\subfigure[]{
		\includegraphics[width=1.0\textwidth]{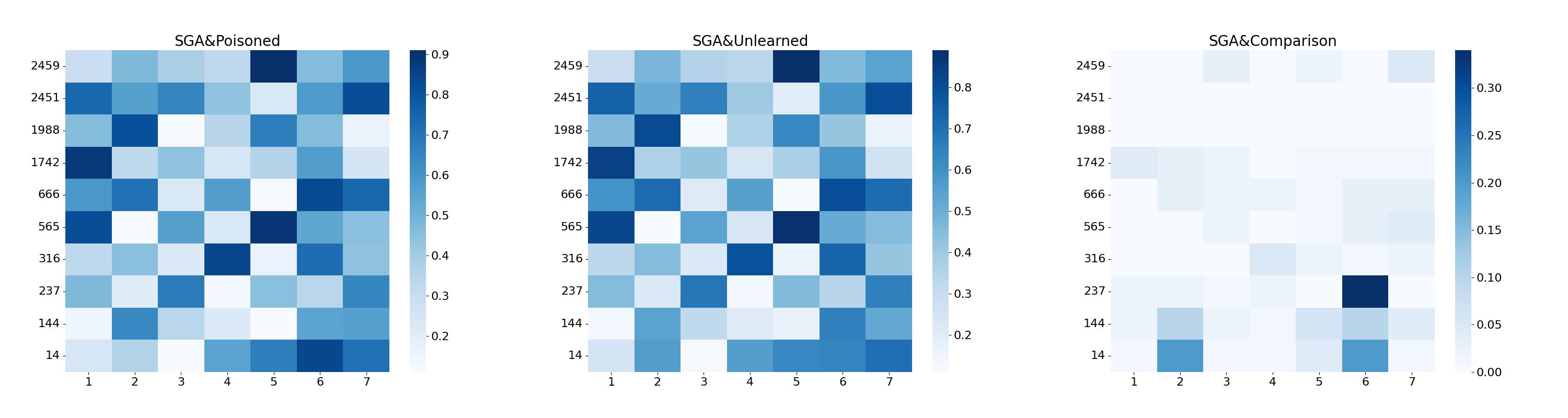}
	}
	\subfigure[]{
		\includegraphics[width=1.0\textwidth]{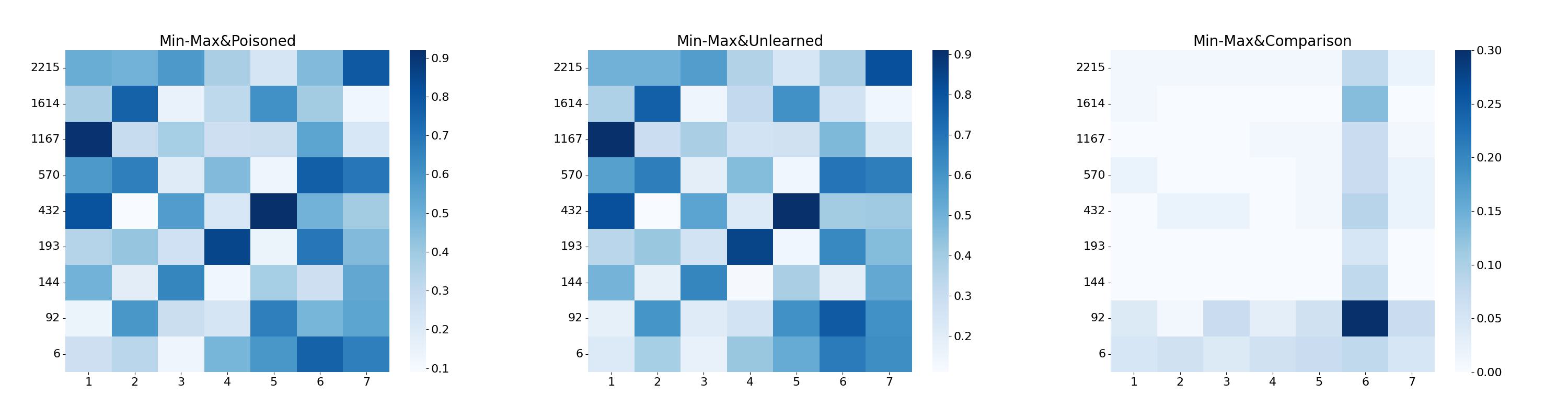}
	}
	\caption{(a) In Cora dataset, heat map of the probability prediction of the poisoned sample with node number 2006 on its surrounding neighboring node after Nettack attacking and after unlearning. (b) In Cora dataset, heat map of the probability prediction of the injected node with node number 2761 on its surrounding neighboring nodes after GANl attacking and after unlearning. (c) In Cora dataset, heat map of probability prediction of nodes with node serial numbers 14 and 237 whoe edges are modifed for all neighboring nodes including them after SGA attacking and after unlearning. (d) In Cora dataset, heat map of the probability prediction of the poisoned sample with node number 570 whose both node feature and
		edge are modifed on its surrounding neighboring nodes after Min-Max attacking and after unlearning.}
	\label{validation}
\end{figure*}
We find that the poisoned GCN is able to be repaired by our method under various attack scenarios, but depending on the level of knowledge about perturbations, there are high and low levels of repair of the poisoned GCN by proposed GraphMU. Specifically, GraphMU has the worst performance in the case of KN-unlearn, and in the other two cases, GraphMU has a relatively good performance but there are some difference. For example, under Nettack and Cora, the performance of GraphMU is more or less the same in both cases of K-unlearn and UK-unlearn. However, under Nettack and Pumbed, the performance of GraphMU in the case of K-unlearn is slightly better than that in the case of UK-unlearn.

To explore the reasons for the performance gap of GraphMU in the above three scenarios, we visualize the distribution of the graph datasets across various scenarios. As shown in Fig. \ref{Visualization}, the visualization of Cora-ML distribution is showcased. The distribution reveals discrepancies between the limited (Fig. \ref{Visualization}(c)) and unlimited (Fig. \ref{Visualization}(d)) detection of perturbations compared to the actual (Fig. \ref{Visualization}(b)). These differences in detected samples, attributed to distinct fine-tuned subgraphs, explain GraphMU's varied repair performance. Despite incomplete detection, the GCN's functionality can be restored since detected samples encompass both adversarial and potentially malicious original graph instances. Utilizing these samples to create fine-tuned subgraph effectively repairs the poisoned GCN.

The average unlearning time of GraphMU on all datasets is shown in TABLE \ref{tab:running time}. The results show that the reparing time is less than the retrain time under four adversarial attack methods and four datasets, which is negligible compared to retraining the poisoned models. The only thing that makes retraining the GCN faster than using GraphMU is under Nettack and Pubmed, which may be related to the proportion of nodes modified by Nettack. Since all four perturbations are proportionally poisoned. Pubmed has a larger amount of data compared to the other three datasets, and thus will be proportionally more poisoned. When constructing the fine-tuned subgraph using our method, the entire graph needs to be traversed. Therefore, the time spent on traversal will be more with Pubmed.

\subsection{The impact of two crucial parameters on model performance}
This section investigates the influence of key parameters on GraphMU. The parameters are the hop number $K$ of the subgraph and the fine-tuning rounds $R$. The hop number $K$ indicates the information richness of the graph data, with higher values indicating more information. The number of fine-tuning rounds $R$ represents the frequency of fine-tuned subgraph application. Experiments were conducted with $K$ values of 2, 3, 4, and 5, as $K=1$ results in scatter plots unsuitable for GCN input, and $K>5$ leads to the original graph, which is the equivalent of retraining.

Fig.\ref{hop-effect} shows the experimental findings on the influence of hop numbers on GraphMU within fine-tuned subgraphs across four adversarial scenarios and datasets. Specifically, we divide the experiment into three scenarios: Known all perturbation information, Known perturbation ratios and Unknown any perturbation information. We find that as the number of hops of the constructed fine-tuned subgraph increases, the repair ability of GraphMU does not change much, and in most cases, constructing a subgraph with two hops is already able to repair poisoned GCN very well.

Experiments were conducted to evaluate the impact of varying fine-tuning rounds on GraphMU performance, with the hop number fixed at 2 and fine-tuning rounds set to 5, 10, 20, and 50, as shown in Fig.\ref{round-effect}. Three different scenarios are also set up here as in the discussion above on the impact of subgraph hopping on model repair performance. we find that as the number of fine-tuning rounds increases, the repair ability of GraphMU gradually decreases under the Cora and CiteSeer datasets, and is even lower than the performance of the poisoned model, which does not change much under the Cora-ML and Pubmed datasets. We conjecture that the complexity of the dataset is the main reason for the effect. GNN trained with complicated dataset is less susceptible to the effects of fine-tuning rounds.

\subsection{The analysis of effectiveness}
In this section, we evaluate the effectiveness of GraphMU by analyzing the impact reduction of perturbations on the overall graph. Experiments were conducted with Nettack, GANI, SGA, and Min-Max attacks, and results are presented in Figure \ref{validation}. The left side illustrates the poisoned GNN's prediction, the middle shows the prediction of graph model after repairing, and the right indicates the prediction difference. We find that the unlearning process resulted in altered prediction probabilities for anomalous nodes' surrounding neighboring nodes, demonstrating GraphMU's efficacy in reducing the poisoned node's influence.

\section{Conclusion}
In this paper, we for the first time explore the problem of GNNs repair and introduce GraphMU which is model-agnostic to repair poisoned GNNs. To demonstrate the effectiveness of our method, we conduct experiments with plenty of adversarial attacks methods and several benchmark datasets in all scenarios. Through extensive experiments, the results show that the performance of the poisoned GCN can be restored with our method. In addition, we have also explored in depth the effects of the number of hops of fine-tuned subgraph construction and the number of fine-tuning rounds on the model performance restoration. The results show that the effect of the number of hops for fine-tuned subgraph construction on GNNs repair is not significant, and the number of rounds of fine-tuning becomes a key factor. Specifically, the best performance of GraphMU is achieved under five rounds of fine-tuning. Meanwhile, as the number of rounds of fine-tuning is further increased, the repair performance of GraphMU will decrease due to overfitting. This finding is important for guiding the restoration practice of the poisoned GNNs.

\section*{Acknowledgment}
This work was partially supported by the National Natural Science Foundation of China under Grant Nos. 62376047 and 62106030 and the Chongqing Education Commission Science and Technology Research Program Key Project under Grant No. KJZD-K202300603, Chongqing Technology Innovation and Application Development Project under Grant No. CSTB2022TIAD-GPX0014, and the Open Foundation of Yunnan Key Laboratory of Software Engineering under Grant No.2023SE204.

\bibliographystyle{IEEEtran}
\bibliography{reference.bib}

\newpage

\vspace{11pt}

\vfill

\end{document}